\documentclass[preprint,showpacs,amsmath]{revtex4-1}
\usepackage{amsmath}
\usepackage{txfonts}
\usepackage{hyperref}
\usepackage{graphicx}
\usepackage{float}
\usepackage{color}
\usepackage[normalem]{ulem}
\usepackage{array}
\usepackage{booktabs}
\usepackage{mathtools}
\usepackage{placeins}

\def\printcomments{1}

\def\done{1}
\def\willnotdo{2}
\def\partlydone{3}

\newcommand{\comment}[2]{                       
        \ifx\printcomments\undefined
        \else
                \ifnum#2=\done
                         #1
                \else
                        \ifnum#2=\willnotdo
                                \textcolor{red}{\sout{#1}}
                        \else
                                \ifnum#2=\partlydone
                                        \textcolor{blue}{#1}
                                \else
                                        \textcolor{red}{#1}
                                \fi
                        \fi
                \fi
        \fi
}

\begin{document}


\newcommand{\E}{E}
\newcommand{\Prob}{P}
\newcommand{\defeq}{\vcentcolon=}

\title{Optimal potentials for diffusive search strategies}
\author{\L{}ukasz Ku\'smierz}
\affiliation{RIKEN Brain Science Institute, 2-1 Hirosawa, Wako, Saitama 351-0198, Japan}
\author{Martin Bier}
\affiliation{Department of Physics, East Carolina University, Greenville, North Carolina 27858, USA}
\author{Ewa Gudowska-Nowak}
\affiliation{Marian Smoluchowski Institute of Physics, Jagiellonian University, ul. \L{}ojasiewicza 11, 30-348  Krak\'ow, Poland}
\affiliation{Mark Kac Complex Systems Research Center, Jagiellonian University, Krak\'ow, Poland}

\begin{abstract}
We consider one dimensional diffusive search strategies subjected to external potentials.
The location of a single target is drawn from a given probability density function (PDF) 
$f_G(x)$ and is fixed for each stochastic realization of the process.
We optimize the quality of the search strategy as measured 
by the mean first passage time (MFPT) to the position of the target.
For a symmetric but otherwise arbitrary distribution $f_G(x)$ we
find the optimal potential that minimizes the MFPT. 
The minimal MFPT is given by a nonstandard measure of the dispersion, 
which can be related to the cumulative R\'enyi entropy. 
We compare optimal times in this model with optimal times obtained 
for the model of diffusion with stochastic resetting, 
in which the diffusive motion is interrupted by 
intermittent jumps (resets) to the initial position.  
Additionally, we discuss an analogy between our results and 
a so-called square-root principle.
\end{abstract}

\maketitle
\section{Introduction}
A random search process, its duration, energetic cost, and optimization are 
frequently analyzed in various interdisciplinary contexts 
\cite{Oshanin,Shlesinger2006,benichou2011intermittent}, 
ranging from diffusion of regulatory proteins on DNA 
\cite{Badrin,slutsky2004kinetics,eliazar2007searching}, 
foraging patterns of animals 
\cite{viswanathan1996levy,perry1997animal,viswanathan1999optimizing,maarell2002foraging,ramos2004levy,bartumeus2005animal,benichou2005optimal,reynolds2007free,reynolds2007displaced,edwards2007revisiting,reynolds2008deterministic,sims2008scaling,humphries2010environmental,Reynolds2015}, 
and page ranking, graph mining and general optimization techniques used in computer science 
\cite{kirkpatrick1983optimization,bohachevsky1986generalized,tong2006fast,yang2009cuckoo,yang2010firefly}. 
Often, the search process becomes confined by the domains of restricted motion, 
or subject to landscapes with distributed targets. 
The questions which then arise naturally are how long it takes to 
locate a target and how to determine optimal search motion. 
When an unsuccessful random search is broken off and a new search 
is started again at the origin, the process is known as 
a random walk with resetting \cite{majumdar2011resetting}. 
Such random walks have recently attracted significant research attention; 
of particular interest is how the resetting rate 
and the features of the diffusive motion (super- or sub-) 
affect the effectiveness of the search 
\cite{kusmierz2014first,kusmierz2015optimal,Montero}. 

The resetting  mechanism is an interference from outside and, as such, 
it is a nonequilibrium modification of the system. 
Accompanied by diffusion in configurational space, 
search with resetting violates detailed balance and 
leads to a current-carrying nonequilibrium stationary state 
\cite{evans2013optimal} described by the stationary distribution $p_s(x)$. 
The latter can be expressed in terms of 
a Boltzmann weight with an effective potential, 
$p_s(x)\propto \exp(-V_{eff}(x))$. 
One is then tempted to ask the following question: is the search described 
by the (equilibrium) Langevin dynamics on $V_{eff}(x)$ 
just as efficient as a diffusion-with-resetting search?
The following approach, proposed in \cite{evans2013optimal}, has addressed this issue: 
authors assumed that the target position is \textit{fixed} and studied the first-passage 
time problem for a diffusive searcher with stochastic resetting with a finite rate. 
Next, optimal search times were compared to those of the equivalent Langevin process, 
i.e. the Langevin process leading to the same stationary state.
It was shown that diffusion with stochastic resetting gives shorter 
search times than diffusion in an effective potential.

One thus may be prompted to conclude that equilibrium dynamics is worse 
as a diffusive search strategy than stochastic resetting. 
In order to show that this is not necessarily the case we focus on 
a slightly different problem: 
given a \textit{distribution} of possible target positions we separately 
optimize both the diffusion with stochastic resetting and 
the diffusion with an external potential. 
The latter optimization is performed in the space of functions, 
whereas the former has only a parameter (the rate of the resetting $r$) that must be optimized. 
The diffusion coefficient $D$ is fixed and without loss of generality we choose $D=1$.

With the optimization of the potential we choose from a space of functions.  
With this greater flexibility, one may expect that the optimization of the potential 
leads to shorter MFPTs than the mere optimization of the scalar resetting rate. 
However, as we will show below, there are cases in which the resetting scenario is still better, 
i.e. the nonequilibrium stochastic resetting gives shorter 
MFPTs than any possible equilibrium dynamics search.


\section{Problem statement}
\subsection{Model}
A random searcher performs one-dimensional 
overdamped Brownian motion in the potential $U$:
\begin{equation}
\mbox{d}X_t=-U'(X_t)\mathrm{d}t+\sqrt{2} \mathrm{d}W_t,
\end{equation}
with $X_0=0$. 
For each realization of the process there exists one target at position $G$, 
which itself is a random variable with a given PDF $f_G(x)$. 
We introduce the first arrival time
\begin{equation}
T = \inf(t:X_t = G),
\end{equation}
which in our case, since the trajectories are almost surely continuous, 
coincides with the first passage time \cite{chechkin2003first}. 
Our aim is to answer the following questions: 
what is the optimal potential $U^*(x)$ for which the MFPT 
$\langle T \rangle$ is minimal? 
And what is the actual minimum achievable MFPT
\begin{equation}
T^*=\min_{U}\langle T \rangle
\end{equation}
in this setup?
Hereafter we assume that $f_G(x)$ is not concentrated at the origin, 
i.e. there is no $\delta$ function at the origin 
(and so $\Prob(G=0)\defeq \mbox{Prob}(G=0)=0$). 
Our approach can be easily generalized to the cases when $\Prob(G=0)=p_0>0$ 
with the following relation
\begin{equation}
\langle T \rangle = (1-p_0) \langle T_0 \rangle,
\end{equation}
where $T_0$ is defined as $T$ given $G\neq0$, 
i.e. it is the conditional random variable $T|(G\neq0)$.
\subsection{Useful definitions}
We split $f_G(x)$ into two parts:
\begin{equation}
f_G(x)=p H(x) f_+(x) + (1-p) H(-x) f_-(-x),
\label{Eq:split}
\end{equation}
where 
$f_+:\mathbb{R}_{\ge 0}\to \mathbb{R}_{\ge 0}$, $f_-:\mathbb{R}_{> 0}\to \mathbb{R}_{\ge 0}$, $p=\Prob(G\ge0)$, 
and $H(x)$ denotes the Heaviside step function. 
Note that with these definitions the normalization is preserved, 
i.e. $\int\limits_0^{\infty}f_{\pm}(x)\mbox{d}x=1$.

By $F_{\pm}(x)=\int\limits_0^{x} f_{\pm}(x')\mbox{d}x'$ 
we denote one-sided cumulative distribution functions. 
Furthermore, we define functions:
\begin{equation}
Q_{\pm}(x):=\sqrt{1-F_{\pm}(x)},
\end{equation}
\begin{equation}
g_{\pm}(x):=\int\limits_0^{x} Q_{\pm}(x')\mbox{d}x',
\end{equation}
and constants:
\begin{equation}
\rho_{\pm}:=\lim_{x\to\infty}g_{\pm}(x)=\int\limits_0^{\infty} Q_{\pm}(x')\mbox{d}x'.
\end{equation}
For symmetric (even) $f_G(x)$ it is straightforward to show that 
$\rho_+=\rho_-$ and $p=\frac{1}{2}$.
\section{Results}
\subsection{General solution for symmetric $f_G(x)$}
\begin{figure}
\begin{center}
\includegraphics[width = 1.0\linewidth]{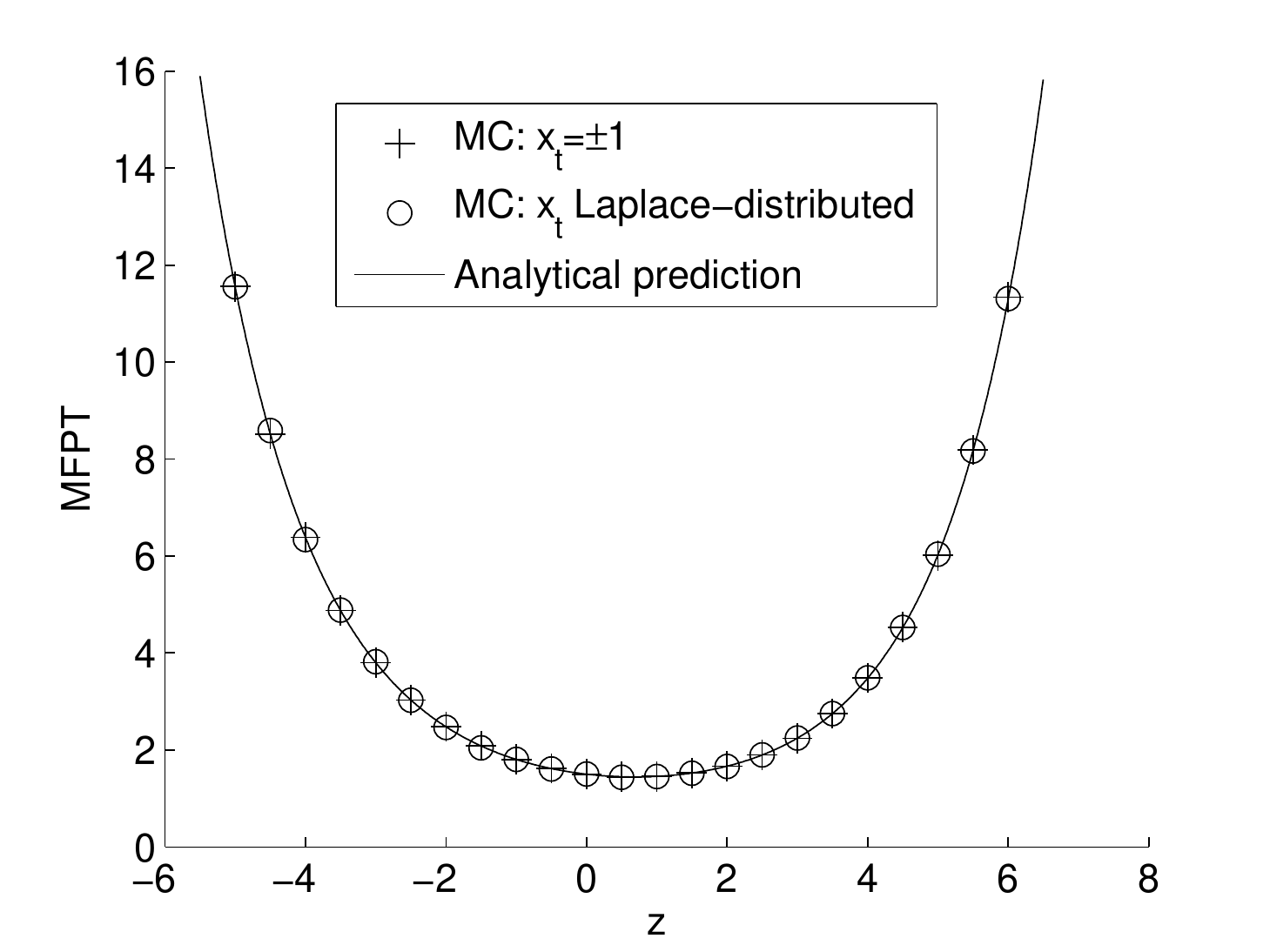}
\caption{A comparison between the analytical prediction 
(\ref{eq:universal-t}) and estimates obtained from $10^6$ 
independent sample trajectories for two different target locations distributions: 
crosses and circles represent results obtained for the two-point distribution 
and for the Laplace distribution, respectively. 
In the latter case the MFPT has been rescaled by the factor 
$\frac{1}{\rho_+^2}=\frac{1}{4}$ to match the universal curve. 
The integration has been performed by means of 
the Euler-Maruyama method with $\Delta t=10^{-3}$, 
with additional correction to avoid the bias of the MFPT estimator, 
as explained in \cite{mannella1999absorbing}. 
Because of the large number of samples, 
error bars would be smaller than or, 
in some cases, comparable to the markers and thus are not included in the plot.}
\label{fig:check-mc}
\end{center}
\end{figure}

In the case of a symmetric $f_G(x)$ we can solve the problem exactly 
(for derivation, see Appendix \ref{app:derivation}). 
The optimal potential reads
\begin{equation}
U^*(x)=- \frac{g_+(|x|)}{\rho_+}\ln{2} -\ln{Q_+(|x|)},
\end{equation}
and the optimal MFPT reads
\begin{equation}
T^*=\frac{\rho_+^2}{\ln{2}}=\frac{1}{\ln{2}}\left(\int\limits_0^{\infty} \sqrt{1-F_+(x')}\mbox{d}x'\right)^2.
\label{eq:optimal-time}
\end{equation}
Moreover, for the class of potentials parametrized by an auxiliary variable $z$: 
\begin{equation}
U_z(x)=- \frac{g_+(|x|)}{\rho_+} z -\ln{Q_+(|x|)}
\label{general_solution_sym_U}
\end{equation}
we observe a universal behavior of the MFPT
\begin{equation}
\langle T_z \rangle = \rho_+^2 \frac{e^{z}+2 e^{-z} + z -3}{z^2}.
\label{eq:universal-t}
\end{equation}
Note that in that case the dependence of the MFPT on $f_G(x)$ comes 
only from the scaling factor. 
We emphasize that the universal expression (\ref{eq:universal-t}) 
is exact and does not involve any approximations. 
Indeed, a perfect agreement between the predicted averages, 
cf.\ Eq.~(\ref{eq:universal-t}), and the averages calculated from $10^6$ 
sample trajectories by means of stochastic simulations is observed 
(see Fig.~\ref{fig:check-mc}). 

%

\subsection{Square-root principle}
The formula for the optimal time (\ref{eq:optimal-time}) is mindful of  
the so-called square-root principle. 
This principle emerges when we consider a simpler, discrete equivalent of a random search.
Assume there are $N$ states (positions) in which we look for exactly one hidden target. 
Let $p_i$ be a probability that $i$-th state is the target. 
The search is performed by randomly sampling positions where the probability of 
picking a position $i$ is $q_i$. 
The values of $q_i$ cannot be changed in the course of the search. 
The question that we want to address is what $q_i$ optimizes the expected number 
of trials before finding the target. 
A straightforward expression can be derived for the optimal $q_i$'s
\begin{equation}
q^*_i=\frac{1}{C}\sqrt{p_i}=\frac{\sqrt{p_i}}{\sum\limits_{j=1}^{N}\sqrt{p_j}},
\end{equation}
i.e. the optimal probability is obtained by taking a square root 
of the probability $p_i$ with a proper normalizing factor $C$. 
Furthermore, the square of the normalizing factor gives the optimum 
for the expected number of trials $\langle n_d\rangle^*$:
\begin{equation}
\langle n_d\rangle^*=C^2=\left( \sum\limits_{j=1}^{N}\sqrt{p_j} \right)^2.
\label{eq:opt-number-trials}
\end{equation}
This natural and simple problem has been considered in the engineering community, 
e.g. in the context of a scheduling data broadcast \cite{vaidya1999scheduling} 
and a replication strategy in peer-to-peer networks \cite{cohen2002replication}.

In the case of a continuous space of possible target locations the problem is 
a little bit more subtle and has been analyzed in \cite{snider2011optimal}. 
In the limit of a small single trial range the sum in 
Eq.\ (\ref{eq:opt-number-trials}) is substituted by an integral and the 
probability $p_i$ by a probability density function $f(x)$, leading to
\begin{equation}
\langle n_c\rangle^*=C^2=\left( \int\limits_{\mathbb{R}}\sqrt{f(x)}\mbox{ d}x \right)^2
\label{eq:opt-number-trials-cont},
\end{equation}
in analogy to Eq.\ (\ref{eq:optimal-time}). 
Note that, in contrast to the problem that we discuss in this paper, 
the square-root principle has been derived for a nonlocal search in discrete time, 
i.e. each trial is independent from previous trials. 
In the case of diffusive search only local moves to close sites are possible, 
i.e. only the neighborhood of the last trial has significant probability of being visited. 
This obviously leads to different formulas for the optimal time 
(or, analogously, the optimal number of trials). 
Nonetheless, the similarity between Eqs.\ 
(\ref{eq:optimal-time}) and (\ref{eq:opt-number-trials-cont}) suggests 
that the square-root principle may be fundamental in random search processes.

\subsection{Properties of $\rho_+$}
Let $G_+$ denote the conditional random variable $G|G\geq0$. 
Its probability distribution function and cumulative distribution function 
are given by $f_+(x)$ and $F_+(x)$ respectively. 
It is easy to see that $\rho_+$ has the same unit as $G_+$. 
How does it relate to moments of $G_+$? 
As $F_{+}(x)$ is non-negative and not larger than one, we have:
\begin{equation}
\rho_+=
\int\limits_0^{\infty} \sqrt{1-F_+(x)}\mbox{d}x 
\geq \int\limits_0^{\infty} \left(1-F_+(x)\right)\mbox{d}x
=\langle G_+ \rangle.
\label{eq:inequality}
\end{equation}
We next describe the large $x$ behavior of $F_+(x)$ with a power law. 
More specifically, we let $\left|F(x)-(1-\frac{C}{x^{\alpha}})\right|\leq \epsilon$ for $x>x_0$. 
This leads to
\begin{equation}
\rho_+ \approx 
\int\limits_0^{x_0} \sqrt{1-F_+(x)}\mbox{d}x+\int\limits_{x_0}^{\infty} \sqrt{\frac{C}{x^{\alpha}}}\mbox{d}x.
\end{equation}
The integral $\int\limits_{x_0}^{\infty} x^{-\frac{\alpha}{2}}\mbox{d}x$ 
is finite if and only if $\alpha>2$, 
which coincides with the condition for the variance to exist. 
We therefore conclude that the finiteness of $\rho_+$ is equivalent to 
the finiteness of the variance:
\begin{equation}
\rho_+ < \infty \iff \langle G_+^2 \rangle -\langle G_+ \rangle^2<\infty.
\label{eq:rho-variance}
\end{equation}

The dispersion properties of $G$ can also be discussed in terms of entropies. 
The entropy of a system is commonly understood as proportional 
to the logarithm of the available phase space.  
However, entropy can also be viewed as the amount of information 
that is associated with the outcome of a measurement.  
Different setups often require different mathematical formalisms 
and different definitions for the entropy have thus been formulated and used   
\cite{shannon1949communication,tsallis1988possible,pennini1998renyi,abe2001heat,renyi1959dimension,zyczkowski2003renyi,wehner2010entropic}.
For our case the quantity $\rho_+$ can be expressed in terms of a, 
recently proposed, cumulative R\'enyi entropy (CRE) $\gamma_{\beta}$. 
For a non-negative random variable $X$ this entropy 
is defined as follows \cite{sunoj2012dynamic}
\begin{equation}
\gamma_\beta(X)=\frac{1}{1-\beta}\log\left(\int\limits_0^{\infty}\bar{F}_{X}^{\beta}(x)\mbox{d}x\right),
\end{equation}
where $\bar{F}_X(x)=1-F_X(x)$ is called the survival function. 
With this definition, the optimal MFPT (in the symmetric case) reads
\begin{equation}
T^*=
\frac{1}{\ln 2} \exp \left[ \gamma_{1/2} \left( G_+ \right) \right]
.
\end{equation}
\comment{
We see that the optimal MFPT is a monotonic function of the cumulative R\'enyi entropy for $\beta=1/2$. 
So, given a PDF of $G$, the CRE quantifies the level of difficulty for the random diffusive search. 
}{\done}
At this point it is convenient to define a new quantity 
\begin{equation}
\bar{S}_q(X)=\frac{1}{1-q}\left(\int\limits_0^{\infty}\bar{F}_X^{q}(x)\mbox{d}x -\langle X \rangle \right),
\end{equation}
which is related to the Tsallis entropy 
\cite{renyi1959dimension,wehner2010entropic,zyczkowski2003renyi} 
in a similar way as the cumulative R\'enyi entropy is related to the R\'enyi entropy. 
We will thus call it the cumulative Tsallis entropy (CTE). 
The CTE is non-negative and for $q\to1$ reduces to 
the cumulative residual entropy \cite{rao2004cumulative}
\begin{equation}
 \bar{S}_1(X)=-\int\limits_0^{\infty}\bar{F}_X(x)\log{\left(\bar{F}_X(x)\right)}\mbox{d}x.
\end{equation}
In our case the CTE can be used to express $\rho_+$ as a sum of two components
\begin{equation}
\rho_+ = \langle X_+ \rangle + \frac{1}{2}\bar{S}_{1/2}(X_+).
\label{eq:rho-with-cte}
\end{equation}
\comment{
The interpretation of Eq.\ (\ref{eq:rho-with-cte}) is straightforward. 
The minimal achievable MFPT by a random search for a given distribution of $G$ is 
related to two features of the distribution: 
the expected distance to the target and a dispersion of the distances with respect to the expected distance. 
The latter is quantified by the CTE. 
}{\done}
It seems that the CTE may serve as a new measure of dispersion, or randomness, 
of a probability distribution and it would be desirable to research 
its properties in detail. 
Also, its generalization to multivariate random variables may be useful. 
This observation will be the subject of another work.

\subsection{Special cases}
In this part we analyze four special cases of a symmetric $f_G(x)$. 
We calculate the optimal potentials and the corresponding optimal 
search times and we discuss their properties.

\subsubsection{Symmetric two-point distribution}
This is the simplest nontrivial symmetric probability distribution. 
The target is at position $x_0$ or at position $-x_0$ with the same probability, i.e. 
\begin{equation}
f_G(x)=\frac{1}{2} \delta(x+x_0) + \frac{1}{2} \delta(x-x_0),
\end{equation}
with $x_0>0$. 
It is easy to check that $f_+(x)=\delta(x-x_0)$ 
and that the CDF is given by the Heaviside function $F_+(x)=H(x-x_0)$. 
The optimal potential is given by the formula
\begin{equation}
\begin{split}
U^*(x)&=
\begin{cases}
-\frac{\ln{2}}{x_0}|x| &\mbox{for } |x|<x_0 \\
\hfil \infty & \mbox{for } |x|\geq x_0
\end{cases}
\end{split},
\end{equation}
which has been plotted in Fig.\ \ref{fig:opt-pot-all}(a). 
The slope $\ln{\frac{2}{x_0}}$ represents a compromise.  
On the one hand, a steeper slope will more rapidly drive the searching particle 
from the initial $x=0$ to a position where there is a probability $\frac{1}{2}$ 
of finding the target. But if the target is not there, then you want to quickly 
reach the other possible position. 
Thermal activation has to get the searching particle back over the barrier at 
$x=0$ and too steep a slope will delay such barrier crossing. 
The infinite potential barrier is also quite easy to understand: 
since the target is either at the position $x_0$ or $-x_0$, 
searching outside the interval $[-x_0,x_0]$ is a waste 
of time. The optimal time reads:
\begin{equation}
T^*=\frac{x_0^2}{\ln{2}}.
\end{equation}
Intuitively, this distribution is ``easy'', because there are only two possible positions 
of the target. More precisely, this distribution should lead to the shortest 
possible MFPT for a given expected distance 
$\langle G_+ \rangle$. 
That this is the case we can verify by inspecting the form of $\rho_+$:
\begin{equation}
\rho_+=x_0= \langle G_+\rangle.
\end{equation} 
\comment{
In relation to Eq.\ (\ref{eq:inequality}), we see that this case actually leads to equality.
At the same time we see that the CTE for this case reads $\bar{S}_{1/2}(G_+)=0$, 
i.e. the two-point distribution is the least dispersed distribution. 
No other distribution with the same 
$\langle G_+ \rangle$ 
can give rise to a shorter MFPT.
}{\done}

\subsubsection{Symmetric uniform distribution}
The uniform distribution is the simplest guess in bounded environments. 
We parametrize it by the expected distance to the target $\lambda=\langle G_+ \rangle$:
\begin{equation}
f_G(x)=
\begin{cases}
\frac{1}{4\lambda}&\mbox{for } |x| \leq 2\lambda\\
\hfil0&\mbox{ for } |x|>2\lambda
\end{cases}.
\end{equation}
%
Straightforward calculations lead to $\rho_+=\frac{4 \lambda}{3}$ 
and the following form of the optimal potential
\begin{equation}
U^*(x)=
\begin{cases}
\left(\left( 1-\frac{|x|}{2\lambda}\right)^{\frac{3}{2}} -1 \right)\ln{2} -\frac{1}{2}\ln{(1-\frac{|x|}{2\lambda})}&\mbox{ for } |x| \leq 2\lambda\\
\hfil \infty &\mbox{ for } |x|>2\lambda.
\end{cases}
\end{equation}
Due to the compact support of $f_G(x)$ there is also an infinite potential well, 
but not as sharp as in the case of two-point distribution (cf.\ Fig.\ref{fig:opt-pot-all}(b)). 
The optimal time is given by the formula
\begin{equation}
T^*=\frac{16\lambda^2}{9\ln2},
\end{equation}
and it is almost twice the corresponding value for the two-point distribution.

\subsubsection{Laplace distribution}
The Laplace distribution maximizes the Shannon entropy for a given $\lambda=\langle G_+ \rangle$, 
thus it is a natural choice if the only information about the distribution 
we have is $\langle G_+ \rangle$. Its PDF has the form
\begin{equation}
f_G(x)=\frac{1}{2 \lambda} e^{-\frac{|x|}{\lambda}},
\end{equation}
from which we easily calculate the optimal potential 
(see Fig.\ \ref{fig:opt-pot-all}(c))
\begin{equation}
U^*(x)=\left(e^{-\frac{|x|}{2 \lambda}}-1\right)\ln{2}+\frac{|x|}{2 \lambda},
\end{equation}
and the optimal time
\begin{equation}
T^*=\frac{4 \lambda^2}{\ln{2}}, 
\label{eq:opt-t-laplace}
\end{equation}
which is exactly four times larger than the corresponding value for the two-point 
distribution. Although the Laplace distribution maximizes the Shannon entropy, it does 
not maximize the MFPT for a given $\langle G_+ \rangle$, as seen from the next example.

\subsubsection{Power-law distribution}
\begin{figure}
\begin{center}
\includegraphics[width = 1.0\linewidth]{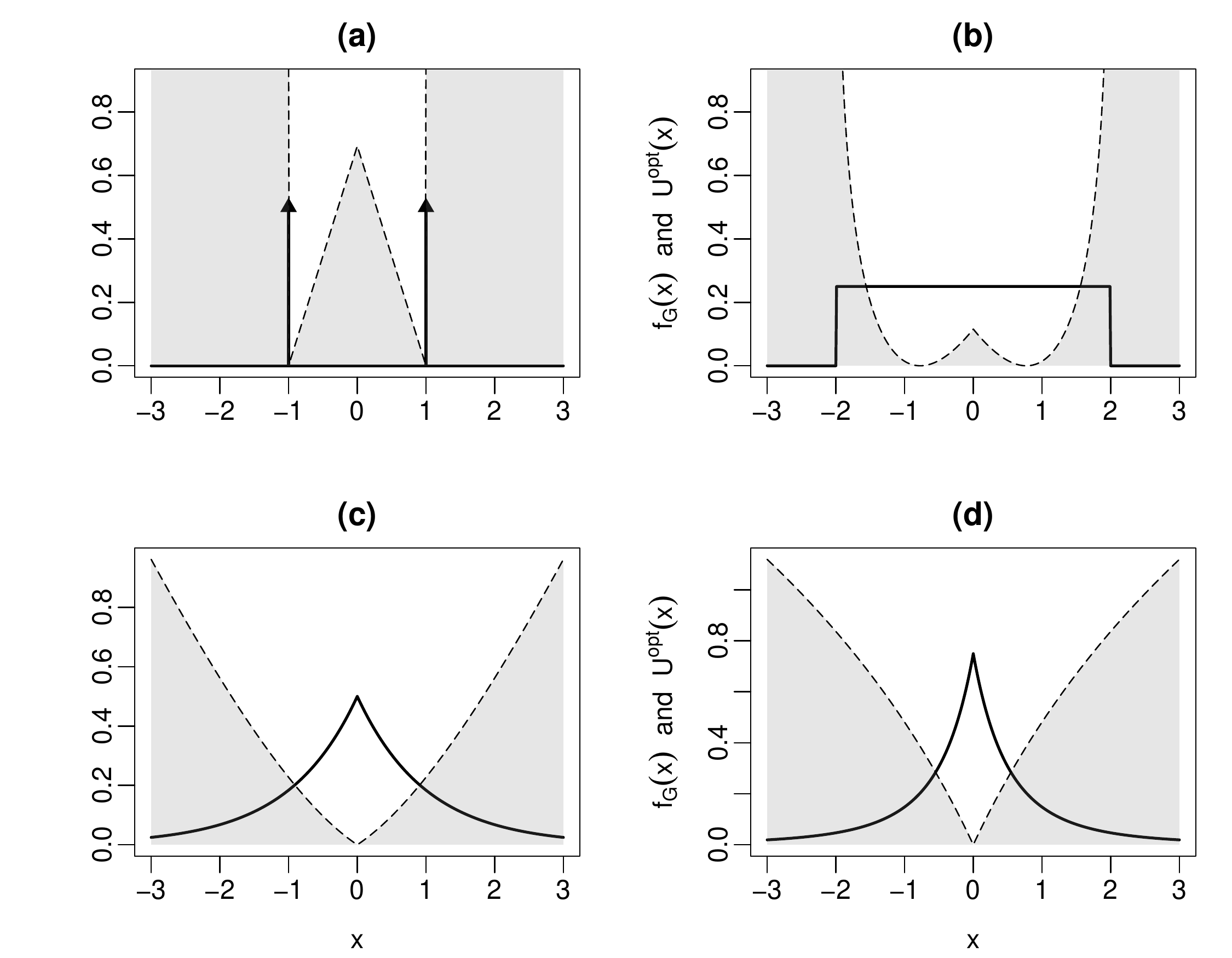}
\caption{Visualizations of the relation between the optimal potential (dashed line 
and shaded area under the curve) and the distribution of a distance to the 
target $f_G(x)$ (solid line). For readability the optimal potentials have been shifted 
so that their minimum values are zero. (a) two point distribution, (b) uniform 
distribution, (c) Laplace distribution, (d) power-law distribution with $\epsilon=2$ 
and $\mu=3$.}
\label{fig:opt-pot-all}
\end{center}
\end{figure}

Our last example is a power-law distribution of the following form
\begin{equation}
f_G(x)=
\frac{\mu \epsilon^{\mu}}{2 (|x|+\epsilon)^{\mu+1}},
\end{equation}
with $\mu>0$ and $\epsilon>0$. 
Parameters $\mu$ and $\epsilon$ determine the tail behavior and the scale, respectively. 
The optimal potential, given by the formula
\begin{equation}
U^*(x)=\frac{\mu}{2}\ln{\left(1+\frac{|x|}{\epsilon}\right)}+ \left(\frac{\epsilon}{\epsilon+|x|} \right)^{\frac{\mu}{2}-1}, 
\end{equation}
grows only logarithmically for large $|x|$, see Fig.\ \ref{fig:opt-pot-all}(d). 
The optimal time reads
\begin{equation}
T^*=\frac{4 \epsilon^2}{(\mu-2)^2\ln2}
=\left(\frac{\mu-1}{\mu-2}\right)^2\frac{4 \lambda^2}{\ln2},
\label{eq:opt-t-powerlaw}
\end{equation}
where $\lambda\equiv \langle G_+ \rangle$, as before, 
denotes the expected value of the 
positive part of the distance to the target distribution.

Note that in the limit $\mu\to\infty$ with $\epsilon=\lambda (\mu-1)$ the power 
law distribution asymptotically approaches the Laplace distribution.
It is easy to check that in this limit the results of the 
Laplace distribution are recovered. 
Comparing Eqs.\ (\ref{eq:opt-t-powerlaw}) and (\ref{eq:opt-t-laplace}) 
we see that, for a fixed $\lambda$, the power-law distribution with any $\mu$ 
leads to a higher value of the optimal MFPT than the Laplace distribution. 
Indeed, it is to be expected, since the heavy tails of the distribution should make 
the search more difficult.

\subsection{Comparison with diffusion with stochastic resetting}
In this section we compare optimal MFPTs of the diffusion with stochastic 
resetting ($T^*_r$) and of the diffusion in a potential ($T^*_U$). The MFPT of 
the diffusion with stochastic resetting with a fixed position of the target ($x$) 
has been calculated in \cite{majumdar2011resetting} and reads:
\begin{equation}
\langle T_r(x) \rangle = \frac{1}{r}\left(e^{\sqrt{r} |x|} - 1 \right),
\end{equation}
where $r$ represents the resetting rate. 
We average this expression over a distribution of possible target positions
\begin{equation}
\langle T_r[f_G] \rangle = \int\limits_{-\infty}^{\infty} f_G(x) \langle T(x) \rangle \mathrm{d} x . 
\end{equation}
For symmetric distributions this leads to
\begin{equation}
\langle T_r[f_+] \rangle = \frac{\tilde{f}_+\left(-\sqrt{r}\right)-1 }{r},
\end{equation}
where  $\tilde{f}(s)$ stands for the Laplace transform of $f(x)$. 
The optimal MFPT, $T^*_r$, is obtained by finding the minimum of 
$\langle T_r[f_+] \rangle $ as a function of $r$. 
Results for different distributions have been summarized in Table 
\ref{tab:opt-comp}. 
It turns out that the resetting performs better for the uniform distribution and 
for the Laplace distributions. 
On the other hand, Langevin dynamics on a potential appears to be the better 
strategy for the two-point distribution and for the power-law distribution. 
Given a distribution, no simple criterion for which of the two search 
strategies is optimal could be formulated. 
With our methods, the ad hoc approach with full derivations is the only one.

\begin{table}
{
\renewcommand{\arraystretch}{1.5}
\begin{tabular}{|c||cc|c|c|}
  \hline 
  $f_G$ & optimized function & $T_r^*$ & $>=<$ & $T_U^*$ \\
  \hline
  two-point & $f_1(z) = \frac{e^z-1}{z^2}$ & $f_1(z^*)\approx 1.54$ & $>$ 
  & $ \frac{1}{\ln 2} \approx 1.44 $ 
  \\
  \hline
  uniform & $f_2(z) = \frac{e^z-1-z}{z^3}$ & $f_2(z^*)\approx 2.19 $ 
  & $<$ & $\frac{16}{9 \ln 2} \approx 2.56$ 
  \\
  \hline
  Laplace & $f_3(z) = \frac{1}{z-z^2}$ & $f_3(z^*)=4$ & $<$ & $\frac{4}{\ln 2}\approx 5.77$ \\
  \hline
  power-law & - & $\infty$ & $\geq$ & $ \left(\frac{\mu-1}{\mu-2}\right)^2\frac{4}{\ln2}$ 
  \\
  \hline
\end{tabular} 
}
\caption{A comparison of $T_r^*$ and $ T_U^*$ for four different PDFs of target locations. 
Distributions are normalized, i.e. $\langle |X| \rangle =\langle X_+ \rangle=1$ 
(note that in both models the optimal time scales as $\langle X_+ \rangle^2$). 
The level of difficulty of the search grows from the top to the bottom. 
However, the MFPT grows in a different manner for each model, 
leading to changes in supremacy between them.}
\label{tab:opt-comp}
\end{table}

\FloatBarrier
\section{Conclusions and discussion}
We have derived an expression for the potential which optimizes the search time 
for a single target whose position is distributed according to a symmetric PDF. 
The optimal MFPT is given by a nontrivial measure of the dispersion, 
which can be rewritten in terms of the cumulative R\'enyi entropy or 
the cumulative Tsallis entropy.
We have compared the optimal search times of our diffusive search scenario with 
the optimal search times obtained in a traditional diffusion-with-stochastic-resetting scenario. 
We have shown that whether one or the other is optimal depends nontrivially on $f_G (x)$.

This study raises additional questions and opens up new possibilities 
for further research. 
The introduced cumulative Tsallis entropy appears naturally in the described 
problem but it is not clear how to generalize it to multivariate random variables. 
It would be thus interesting to analyze 
optimal potentials in a multidimensional diffusive search, 
in which the optimal MFPT may involve the desired generalization.

Stochastic resetting has been assumed to take place with the same intensity 
across the whole space. A more general process with $r=r(x)$ could exploit the 
information about $f_G(x)$ more effectively. 
We speculate that, for a given $f_G(x)$, the optimal non-homogeneous resetting 
$r^*(x)$ is always more effective than the search in the optimal potential, 
but a proof is not yet known.

Gaussian diffusion is only one example of random search processes used in
the context of random search strategies.
L\'evy flights and L\'evy walks
\cite{mandelbrot1983fractal,mandelbrot1983fractal,shlesinger1986levy,zaburdaev2015levy}
have been proven to outperform normal diffusion in different setups of random search strategies
\cite{viswanathan1999optimizing,bartumeus2002optimizing,raposo2003dynamical,bartumeus2005animal}.
A search with L\'evy flights and resetting could be a superior strategy in both discrete 
\cite{kusmierz2014first} and continuous \cite{kusmierz2015optimal} time.  
How L\'evy flights perform in the framework of optimal potentials is also still to be determined.

\comment{
We have compared the MFPTs of diffusion with stochastic resetting to the MFPTs of diffusion in a potential. 
Of course, resetting and a potential can also be used in conjunction.
Stationary distributions of a Brownian particle with stochastic resetting in 
a potential landscape have been studied \cite{pal2015diffusion}, 
but first passage times in such systems are yet to be explored.
Surely the combination of a potential and resetting can lead to better search efficiency. 
Indeed, it can be shown that the MFPT of such a system has the global infimum at $0$, 
if the optimization of $r$ and $U$ is performed jointly. 
This is easy to understand: 
the optimal way to perform a search in this case is to run as fast as possible 
in one of the possible directions, and reset from time to time. 
This corresponds to a steep potential with the maximum at the initial position. 
By changing the steepness of the potential and $r$ together one can achieve 
arbitrarily short MFPT. 
In practice, however, energy constraints would limit the feasibility of such a search. 
}{\done}

\comment{
Recently it was proven \cite{pal2017first} 
that optimal sharp (i.e.\ deterministic and periodic) resetting leads to shorter MFPTs 
than optimal stochastic resetting. 
Could it be that it is also better than any potential? 
Indeed, optimal sharp resetting for the two-point distribution $f_G$ gives a shorter MFPT 
($\approx1.34$) than the corresponding values for the optimal stochastic resetting 
and the optimal potential (cf.\ Table \ref{tab:opt-comp}). 
Note, however, that the proof in \cite{pal2017first} 
does not apply to our model with nontrivial distributions of target locations, 
because in \cite{pal2017first} it is assumed that the entire process starts again after each reset.
This corresponds to drawing a new value of $G$ after each reset. 
In contrast, in our scenario the target stays at its location until it is found.
The MFPT of a sharp resetting process for a fixed period of resets scales 
with the distance to the target as $x e^{x^2/2}$, 
thus the MFPT for both power-law and Laplace distributions is infinite. 
}{\done}

\comment{
Nontrivial boundary conditions such as impenetrable walls can change the 
results of the search optimization qualitatively. 
For example, a nonzero stochastic resetting rate is not always advantageous 
for a search in bounded spaces \cite{christou2015diffusion}.
In the so-called narrow escape problem 
\cite{schuss2007narrow,benichou2008narrow,lindsay2015narrow}
it has been shown that the dependence of a mean escape time is a nonmonotonic function 
of the range of interaction with the spherical boundary \cite{grebenkov2017diffusive}.
Note that impenetrable walls are equivalent to infinite potential barriers, 
so they introduce constraints on the potentials used in the optimization. 
}{\done}

\comment{
We have focused here on the first moment of first passage times, 
but higher moments, and especially the variance, may be useful in characterizing search processes. 
For instance, it has been shown that in the case of the optimal stochastic resetting 
of almost any stochastic process (where the target will almost surely be hit in finite time),
the MFPT is equal to the standard deviation of first passage times \cite{reuveni2016optimal}. 
Whether any similar universality of higher moments holds in our scenario 
will be a subject of future investigation. 
}{\done}

\comment{
Higher moments of first passage times are indirectly related to the so-called ``mortal walkers'' 
\cite{yuste2013exploration,abad2013evanescent,campos2015optimal}, 
Mortal walkers have a limited time to arrive at the target.  
In practical applications this could be the case if the searching particle is a nucleus that 
is subject to radioactive decay.  
If the walker ``dies'' before finding the target, we effectively get an infinite MFPT.  
With the lifetime of the walker another characteristic timescale is added to the setup.  
This addition changes the optimal potential for a diffusive search and 
it also changes the optimal rate for stochastic resetting.  
For example, in the case of short lifetimes of mortal walkers one can expect 
the optimal potential for the two-point distribution to be steeper in order to 
allow the mortal walker to arrive at the target within the limited time.  
Mortal walkers could be the subject of another interesting research venue.      
}{\done}

\comment{
``Vicious walkers'' 
\cite{gennes1968soluble,fisher1984walks,guttmann1998vicious,krattenthaler2000vicious,schehr2008exact} 
perform diffusion on a line.  
They are ``vicious'', because when two walkers meet they annihilate each other.  
Survival times of vicious walkers in external potentials have been studied 
\cite{bray2004vicious},  
but to our knowledge there has been no analysis of a collective of searching vicious walkers.  
The general analysis of multiple searching particles that diffuse and interact may have practical 
significance as this is what bacteria are commonly understood to do.  
Such analysis, however, is another subject for possible future study.
}{\done}

\begin{acknowledgments}
This project has been supported in part (\L{}.K. and E.G-N.) by the grant from the National Science Center (2014/13/B/ST2/02014).
\end{acknowledgments}
\appendix

\section{Derivation}
\label{app:derivation}
\subsection{General case}
Here we include a sketch of the derivation. For a given PDF $f_G(x)$, the mean first passage time (MFPT) is calculated from the formula
\begin{equation}
\langle T \rangle = \int\limits_{-\infty}^{\infty}\langle T|G=x_t \rangle f_G(x_t)\mbox{d}x_t.
\end{equation}
Since $\langle T|G=x_t \rangle$ depends on the sign of $x_t$, we split the integral into two parts, using definition (\ref{Eq:split})
\begin{equation}
\langle T \rangle = p \int\limits_{0}^{\infty}\langle T|G=x_t \rangle f_+(x_t)\mbox{d}x_t 
+  (1-p) \int\limits^{0}_{-\infty}\langle T|G=x_t \rangle f_-(-x_t)\mbox{d}x_t.
\label{Eq:mfat2}
\end{equation}
The MFPT to a given point for a particle undergoing brownian motion in a potential 
is given by the formula \cite{gardiner1985handbook}:
\begin{equation}
\langle T|G=x_t \rangle =
\begin{cases}
\int\limits_0^{x_t} \mbox{d}x e^{U(y)} \int\limits_{-\infty}^{y}\mbox{d}x e^{-U(x)}&\mbox{for }x_t\geq0\\
\int\limits^0_{x_t} \mbox{d}x e^{U(y)} \int\limits^{\infty}_{y}\mbox{d}x e^{-U(x)}&\mbox{for }x_t<0
\end{cases}.
\label{Eq:mfat_given_point}
\end{equation}
To proceed, we split the potential into the positive and negative parts $U(x)=H(x)U_+(x)+H(-x)U_-(-x)$, 
and plug (\ref{Eq:mfat_given_point}) into (\ref{Eq:mfat2}). 
After simple algebraic manipulations we arrive at the following expression:
\begin{equation}
\begin{split}
\langle T \rangle = p \int\limits_0^{\infty} \mbox{d}x e^{U_+(x)}Q_+(x)^2\left( \int\limits_0^{\infty}\mbox{d}y e^{-U_-(y)} +\int\limits_0^{x}\mbox{d}y e^{-U_+(y)} \right) +\\
+ \left(1-p\right) \int\limits_0^{\infty} \mbox{d}x e^{U_-(x)}Q_-(x)^2\left( \int\limits_0^{\infty}\mbox{d}y e^{-U_+(y)} +\int\limits_0^{x}\mbox{d}y e^{-U_-(y)} \right)
\end{split},
\label{mfat_nice_expression}
\end{equation}
with $Q_{\pm}(x)^2=1-F_{\pm}(x)$. 
In the next step we treat the MFPT as a functional of $U_+(x)$ and $U_-(x)$, 
and calculate variational derivatives $\frac{\delta \langle T \rangle}{\delta U_+(x_0)}$ and $\frac{\delta \langle T \rangle}{\delta U_-(x_0)}$. 
Equating them to zero leads to the set of integral equations:
\begin{equation}
\begin{cases}
p e^{2 U_+(x)}Q_+(x)^2\left(\int\limits_0^{\infty}\mbox{d}z e^{-U_-(z)}+\int\limits_0^{x}\mbox{d}z e^{-U_+(z)}\right) 
=\left(1-p\right)\int\limits_0^{\infty}\mbox{d}z e^{U_-(z)}Q_-(z)^2 +p \int\limits_x^{\infty}\mbox{d}z e^{U_+(z)}Q_+(z)^2
\\
\left(1-p\right)e^{2 U_-(x)}Q_-(x)^2\left(\int\limits_0^{\infty}\mbox{d}z e^{-U_+(z)}+\int\limits_0^{x}\mbox{d}z e^{-U_-(z)}\right) 
=p\int\limits_0^{\infty}\mbox{d}z e^{U_+(z)}Q_+(z)^2 +\left(1-p\right) \int\limits_x^{\infty}\mbox{d}z e^{U_-(z)}Q_-(z)^2
\end{cases}.
\end{equation}
These coupled equations can be rewritten as decoupled differential equations. 
Solving these equations lead to the general solution with three arbitrary constants 
$(z_-,z_+,C_0)$. The optimal potential takes the following form
\begin{equation}
\begin{cases}
U_+(x)= -\frac{g_+(x)}{\rho_+} z_+ -\ln{Q_+(x)} +C_0 \\
U_-(x)= -\frac{g_-(x)}{\rho_-} z_- -\ln{Q_-(x)}
\end{cases}.
\label{asymmetric_opt_potential}
\end{equation}
The potential of the form ($\ref{asymmetric_opt_potential}$) leads to the following 
expression for the MFPT
\begin{equation}
\begin{split}
\langle T_{(z_+,z_-,C_0)} \rangle = p\frac{\rho_+^2}{z_+^2}\left(e^{-z_+}+z_+ - 1\right) +(1-p)\frac{\rho_-^2}{z_-^2}\left(e^{-z_-}+z_- - 1\right) +\\
+\frac{\rho_+ \rho_-}{z_+ z_-}\left( p e^{C_0}(1-e^{-z_+})(e^{z_-}-1) + (1-p) e^{-C_0}(1-e^{-z_-})(e^{z_+}-1)\right)
\end{split}.
\end{equation}

\subsection{Symmetric $f_G(x)$}
As mentioned before, a symmetric target distribution leads to $p=\frac{1}{2}$ and $\rho_+$=$\rho_-$. 
Additionally, for $x>0$, also $Q_+(x)=Q_-(x)$ and $g_+(x)=g_-(x)$ hold. 
This symmetry should bring about symmetry of the optimal potential, which we further assume. 

With these assumptions we obtain the general solution of the form (\ref{general_solution_sym_U}). 
We plug this solution into formula for the MFPT (\ref{mfat_nice_expression}) 
and arrive at the following expression for the MFPT in function of $z$ and $\rho_+$
\begin{equation}
\langle T_z \rangle = \rho_+^2 t(z):= \rho_+^2 \frac{e^{z}+2 e^{-z} + z -3}{z^2}.
\end{equation}
In the last step we find the minimum of $t(z)$, which turns out to admit a simple form
\begin{equation}
\begin{cases}
z^*&=\ln{2}\\
t(z^*)&=\frac{1}{\ln{2}}
\end{cases}.
\end{equation}

\bibliography{citations}
\bibliographystyle{phjcp}
\end{document}